\documentstyle[psfig,times]{mn}  
\begin{document}
\def\reef{\par\noindent\hang}
\def\etal{et al.\ }
\def\eg{{\em eg.\ }}
\def\etc{{\em etc.\ }}
\def\ie{{\em i.e.\ }}

\def\spose#1{\hbox to 0pt{#1\hss}}
\def\approxlt{\mathrel{\spose{\lower 3pt\hbox{$\sim$}}
	\raise 2.0pt\hbox{$<$}}}
\def\approxgt{\mathrel{\spose{\lower 3pt\hbox{$\sim$}}
	\raise 2.0pt\hbox{$>$}}}
	
\def\Mdot{\hbox{$\dot M$}}
\def\degmark{$^\circ$}
\def\<{\thinspace}
\def\s{\hbox{\phantom{5}}}	
\def\ss{\s\s}		
\def\sss{\ss\s}		
\def\ssss{\ss\ss}	
\def\lit{\obeyspaces\obeylines}
%
\def\arc{{\rm\thinspace arcsec}}
\def\cm{{\rm\thinspace cm}}
\def\ct{{\rm\thinspace ct}}
\def\erg{{\rm\thinspace erg}}
\def\eV{{\rm\thinspace eV}}
\def\g{{\rm\thinspace g}}
\def\G{{\rm\thinspace G}}
\def\ga{{\rm\thinspace gauss}}
\def\K{{\rm\thinspace K}}
\def\keV{{\rm\thinspace keV}}
\def\m{{\rm\thinspace m}}
\def\km{{\rm\thinspace km}}
\def\kpc{{\rm\thinspace kpc}}
\def\Lsun{\hbox{$\rm\thinspace L_{\odot}$}}
\def\rad{{\rm\thinspace rad}}
\def\MeV{{\rm\thinspace MeV}}
\def\Mpc{{\rm\thinspace Mpc}}
\def\Msun{\hbox{$\rm\thinspace M_{\odot}$}}
\def\pc{{\rm\thinspace pc}}
\def\ph{{\rm\thinspace photons}}
\def\s{{\rm\thinspace s}}
\def\yr{{\rm\thinspace yr}}
\def\sr{{\rm\thinspace sr}}
\def\Hz{{\rm\thinspace Hz}}
\def\GHz{{\rm\thinspace GHz}}
\def\W{{\rm\thinspace W}}
\def\cmps{\hbox{$\cm\s^{-1}\,$}}
\def\ctps{\hbox{$\ct\s^{-1}\,$}}
\def\ctpsparcsecsq{\hbox{$\ct\s^{-1}\arc^{-2}\,$}}
\def\cmsq{\hbox{$\cm^2\,$}}
\def\cmcu{\hbox{$\cm^3\,$}}
\def\pHz{\hbox{$\Hz^{-1}\,$}}
\def\pcmcu{\hbox{$\cm^{-3}\,$}}
\def\ergcmcups{\hbox{$\erg\cm^3\ps\,$}}
\def\ergpcmps{\hbox{$\erg\cm^{-3}\s^{-1}\,$}}
\def\ergpcmsqps{\hbox{$\erg\cm^{-2}\s^{-1}\,$}}
\def\ergpspcmsq{\hbox{$\erg\cm^{-2}\s^{-1}\,$}}
\def\ergpspkpcsq{\hbox{$\erg\s^{-1}\kpc^{-2}\,$}}
\def\ergpspA{\hbox{$\erg\s^{-1}\AA^{-1}\,$}}
\def\ergpspcmsqpA{\hbox{$\erg\s^{-1}\cm^{-2}$\AA$^{-1}\,$}}
\def\ergpcmsqpspsqarcsec{\hbox{$\erg\cm^{-2}\s^{-1}\arc^{-2}\,$}}
\def\ergpcmsqpspapsqarcsec{\hbox{$\erg\cm^{-2}\s^{-1}\AA^{-1},\arc^{-2}\,$}}
\def\ergps{\hbox{$\erg\s^{-1}\,$}}
\def\gpcm{\hbox{$\g\cm^{-3}\,$}}
\def\gpcmps{\hbox{$\g\cm^{-3}\s^{-1}\,$}}
\def\gps{\hbox{$\g\s^{-1}\,$}}
\def\WpHz{\hbox{$\W\Hz^{-1}\,$}}
\def\kmps{\hbox{$\km\s^{-1}\,$}}
\def\ksec{\hbox{$ksec\,$}}
\def\Lsunppc{\hbox{$\Lsun\pc^{-3}\,$}}
\def\Msunpc{\hbox{$\Msun\pc^{-3}\,$}}
\def\Msunpkpc{\hbox{$\Msun\kpc^{-1}\,$}}
\def\Msunppc{\hbox{$\Msun\pc^{-3}\,$}}
\def\Msunppcpyr{\hbox{$\Msun\pc^{-3}\yr^{-1}\,$}}
\def\Msunpyr{\hbox{$\Msun\yr^{-1}\,$}}
\def\pcm{\hbox{$\cm^{-1}\,$}}
\def\pcmsq{\hbox{$\cm^{-2}\,$}}
\def\kpcsq{\hbox{$\kpc^{2}\,$}}
\def\pcmsqpkeVps{\hbox{$\cm^{-2}\keV^{-1}\s^{-1}\,$}}
\def\pmsq{\hbox{$\m^{-2}\,$}}
\def\radpmsq{\hbox{$\rad\m^{-2}\,$}}
\def\pcmcuK{\hbox{$\cm^{-3}\K$}}
\def\phps{\hbox{$\ph\s^{-1}\,$}}
\def\phpcmsqps{\hbox{$\ph\cm^{-2}\s^{-1}\,$}}
\def\pHz{\hbox{$\Hz^{-1}\,$}}
\def\pMpc{\hbox{$\Mpc^{-1}\,$}}
\def\pMpccu{\hbox{$\Mpc^{-3}\,$}}
\def\MsunpMpccu{\hbox{$\Msun\Mpc^{-3}\,$}}
\def\ps{\hbox{$\s^{-1}\,$}}
\def\psqcm{\hbox{$\cm^{-2}\,$}}
\def\psr{\hbox{$\sr^{-1}\,$}}
\def\pyr{\hbox{$\yr^{-1}\,$}}
\def\kmpspMpc{\hbox{$\kmps\Mpc^{-1}$}}
\def\Msunpyrpkpc{\hbox{$\Msunpyr\kpc^{-1}$}}

\title{ {\sl Chandra} detection of the intracluster medium around
3C294  at $z=1.786$ } 

\author[A.C. Fabian et al ]
{\parbox[]{6.in} {A.C. Fabian, C.S. Crawford, S. Ettori and J.S. Sanders 
\\
\footnotesize
Institute of Astronomy, Madingley Road, Cambridge CB3 0HA \\}}

\maketitle
\begin{abstract}
We present a {\sl Chandra} observation of the powerful radio galaxy
3C294 showing clear evidence for a surrounding intracluster medium. At
a redshift of 1.786 this is the most distant cluster of galaxies yet
detected in X-rays. The radio core is detected as a point source,
which has a spectrum consistent with a heavily-absorbed power law
implying an intrinsic 2-10\keV\ luminosity of $\sim10^{45}$\ergps. A
small excess of emission is associated with the southern radio
hotspots. The soft, diffuse emission from the intracluster medium is
centred on the radio source. It has an hour-glass shape in the N--S
direction, extending to radii of at least $100\kpc$, well beyond the
radio source. The X-ray spectrum of this extended component is fit by
a thermal model with temperature $\sim5\keV$, or by gas cooling from
above 7\keV\ at rates of $\sim400-700$\Msunpyr. The rest-frame
0.3--10\keV\ luminosity of the cluster is
$\sim4.5\times10^{44}$\ergps.The existence of such a cluster is
consistent with a low density universe.

\end{abstract}

\begin{keywords}  
X-rays: galaxies --
galaxies: active: clusters: individual (3C294) --
intergalactic medium --
cosmology: observations

\end{keywords}

\section{Introduction}

Radio galaxies act as luminous beacons which are detectable across the
Universe. Powerful radio galaxies at low redshift, such as Cygnus A,
lie in rich clusters of galaxies and it is possible that more distant
examples too are in dense environments. They may therefore provide the
means to discover the most distant clusters and thus enable the study
of cluster evolution. Here we present the detection of luminous
diffuse X-ray emission surrounding 3C294, at redshift $z=1.786$. This
is 40 per cent higher in redshift than previously reported diffuse 
cluster X-ray emission (Rosati et al 1999).

3C294 is a very powerful FR~II radio source, associated with an
emission-line galaxy. At high resolution the radio structure shows a
Z-shaped (double-hotspot) morphology suggestive of precessing jets
originating from the weak flat-spectrum core (McCarthy et al 1990).
The galaxy is embedded in a luminous Lyman-$\alpha$ halo, elongated to
the North and South, and thus roughly aligned with the radio source
direction (at position angle 20\degmark). The Lyman-$\alpha$ nebula
extends over a $\sim$75$\times$125\kpcsq area (assuming $H_0=50$\kmpspMpc\
and a cosmological deceleration parameter of $q_0=0.5$), and is
brighter on the northern side of the source toward the side of the
closer and brighter radio hotspot. This side of the nebula has a
triangular morphology, such that the radio core is at its southern
apex (McCarthy et al 1990); there is also a hint that the inner part
of the southern side of the nebula mirrors this shape. This
(bi-)conical morphology is similar to an illumination cone caused by
anisotropic radiation from a central ionizing continuum, perhaps due
to dust scattering of radiation from a quasar nucleus. The
Lyman-$\alpha$ emission shows a large velocity shear of
$\sim$1500\kmps across the whole nebula. The higher ionization lines
of CIV$\lambda$1550, CIII]$\lambda$1909 and HeII$\lambda$1640 are also
spatially extended and seem to share this velocity field (McCarthy,
Baum \& Spinrad 1996).

The northern triangular shape can also be seen on a smaller scale in
the K$'$ image of Stockton, Canalizo \& Ridgway (1999); there is no
obvious extension to the south of the radio core in the near-IR. The
K$'$ image suggests that the apex of the cone may be offset to the NW
from the radio core position by a small amount ($\sim0.2$ arcsec). At
the redshift of the radio galaxy, the K$'$ emission is expected to be
dominated by a stellar population, with little contribution from line
emission. The image shows a few resolved knots within the extended
continuum, none of  which are spatially coincident with any of the radio
components.

Benitez, Martinez-Gonzalez \& Gonzalez-Serrano (1995) find a slight
excess of faint R-band objects within a 45 arcsec radius
($\sim380$kpc) region around 3C294, suggestive of the presence of a
poor cluster of galaxies around the radio galaxy. The radio source is
highly depolarized (Liu \& Pooley 1991), indicating a 
dense surrounding medium.

X-ray emission was detected from the direction of 3C294 using archival
data from the {\sl ROSAT} satellite (Crawford \& Fabian 1996).  This
PSPC image had too few counts either to resolve any spatial extent or
discriminate between a thermal or non-thermal origin for the emission.
A subsequent long HRI exposure showed the source to be very faint, and
possibly spatially extended (Hardcastle \& Worrall 1999; Dickinson et
al 1999). Here we present the first images clearly showing extended 
diffuse X-ray emission. 

\section{Observations}
3C294 was observed for 19.5ksec with the {\sl Chandra} X-ray
observatory on 2000 October 29 and processed with CXCDS version
R4CU5UPD11.1. The telescope (Weisskopf et al 2000) was pointed such
that the target appears one arcmin from the centre of chip 7 (S3) in
ACIS-S. It is thus one arcmin from the nominal aimpoint and the
on-axis point-spread function applies. The lightcurve shows no
evidence for particle background flares in the detector during the
observation, and we have extracted the X-ray data from the total
exposure.

\subsection{Images}
\begin{figure}
\psfig{figure=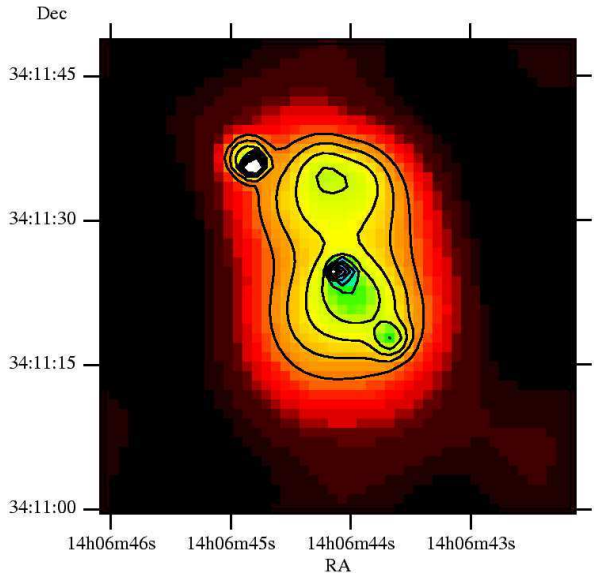,width=0.45\textwidth,angle=0}
\caption{\label{fig:xraycont}
Adaptively smoothed image and contours of the 0.5-5\keV\ {\sl Chandra
} emission from 3C294. The significance level for smoothing is set at
$2\sigma$. Contours shown start at 0.048 ct arcsec$^{-2}$, doubling at
each level. 1 arcsec corresponds to 8.4kpc for the assumed cosmology.
}
\end{figure}

\begin{figure}
\psfig{figure=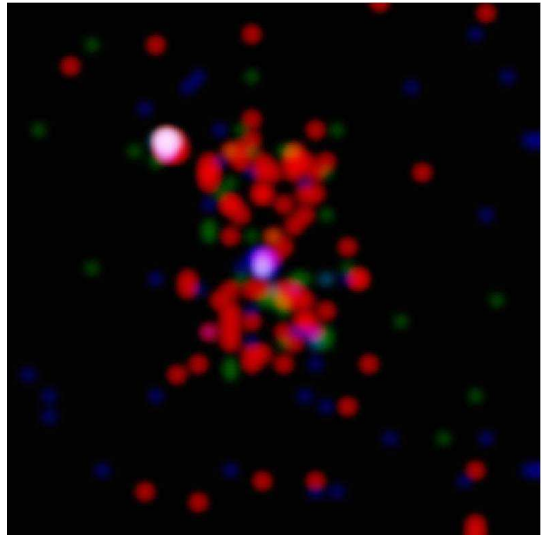,width=0.4\textwidth,angle=0}
\caption{\label{fig:xraycol}
X-ray colour map of the {\sl Chandra } emission from 3C294; 0.5-1\keV\
is shown as red, 1-2\keV\ as green, and 2-5\keV\ as blue. The data
have been binned by 2 pixels (i.e. one arcsec), and smoothed by 2
pixels. North is to the top, East to the left, and the image is
approximately 1.2~arcmin on a side. }
\end{figure}

Inspection of the X-ray image shows a number of immediately apparent
features (Figs~\ref{fig:xraycont}, \ref{fig:xraycol}). The weak radio
core shows up as the brightest X-ray feature, a point source with hard
X-ray colours. The outer southern radio hotspot (feature H$_S$ in the
McCarthy et al map) is spatially coincident with an X-ray excess seen
in both hard and soft bands. This excess is only about $2\sigma$ above
the surrounding diffuse emission. There is a slight hint that there is
also excess X-ray emission at the positions of the inner hotspot and
lobe (K$_S$ and L$_S$) on this side (Fig~\ref{fig:xraycont}). We find
no evidence for any X-ray emission associated with the hotspots to the
northern side of the radio source. Curiously though, there is an X-ray
point source at RA=14:06:44.810 Dec=+34:11:35.74 (J2000) that
continues the line of the radio source axis on this side. At a
distance of $\sim$ 15\ arcsec from the core, the source is too far away
to be associated with any known radio components. It is probably a
serendipitous background source or even another active galaxy within
the cluster, but we find no evidence for a counterpart in the
Digitized Sky Survey or any archival optical images of this field.

The bright X-ray source associated with the radio core clearly lies at
the centre of a soft and spatially extended component of X-ray
emission (Fig~\ref{fig:xraycol}). This very extended component shares
and extends the biconical structure of the Lyman-$\alpha$ nebula; a
triangular shape to the north, with an opening angle of around
80\degmark, and one to the south with a slightly larger opening angle,
both with the radio core at an apex. The total extended shape
resembles an hourglass. 

\begin{figure}
\psfig{figure=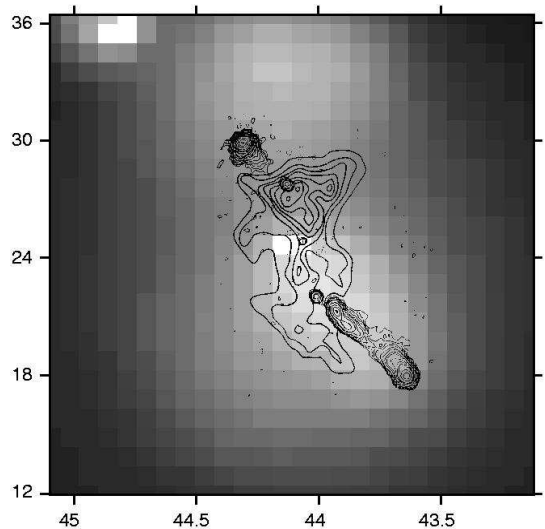,width=0.4\textwidth,angle=0}
\caption{\label{fig:radioandxray}
Greyscale smoothed image of {\sl Chandra } emission from 3C294
(Fig.~1), with the radio (tightly bunched contours) and Ly$-\alpha$
(more open contours) emission superposed (taken from McCarthy et al
1990).}
\end{figure}

The two sides of the structure are much more evenly matched in flux
than is seen in the Lyman-$\alpha$ or near-IR. If anything, the
southern side is slightly brighter, particularly close in to the radio
core, whereas the northern side appears to have a slight deficit in
the cone near to the core (Fig~\ref{fig:xraycont}).  We cannot,
however, rule out the possibility that the inner radio hotspot and
lobe contribute some of this excess brightness close to the core in
the southern cone. The soft component extends out to radii of $\sim$13
arcsec (just over 100kpc) to both the north and south, almost twice as
far as the extent of the radio source structure.

\subsection{Spectra}

\begin{figure}
\psfig{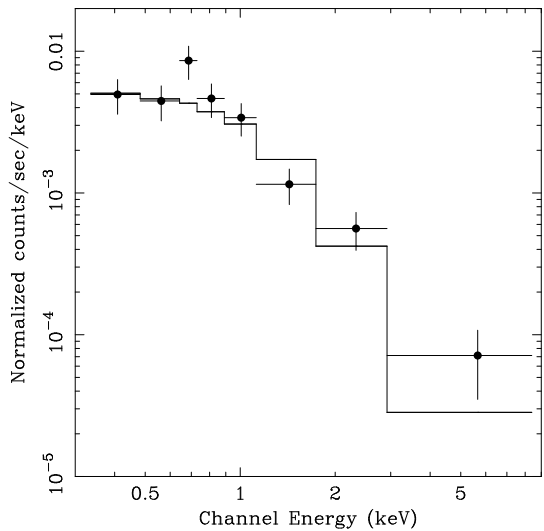}
\caption{\label{fig:xraysp} Spectrum of the soft extended component of X-ray emission
(solid circle markers), with best-fit isothermal 
model of temperature 4.9\keV. 
}\end{figure}

We extracted the spectrum of the soft extended emission in both the
northern and southern cones, excluding a small region encompassing the
central bright core and the bright source $\sim$15arcsec to the NW.
The data have been grouped into bins with a minimum of 15 counts. We
fit the spectrum with a thermal MEKAL model, although with only about
110 counts from the source we are not able to place strong constraints
on the emission properties. We assume that the extended emission is at
the redshift of the radio galaxy, that there is no absorption in
excess of the line-of-sight Galactic hydrogen column of
$1.2\times10^{20}$\pcmsq, and freeze the abundance to be
0.3$\times$~Solar. The best fit (reduced-$\chi^2$ of 1.58 for 6
degrees of freedom) yields a temperature for an isothermal spectrum of
$kT_X=5.0^{+2.6}_{-1.5}$\keV\ (Fig\ref{fig:xraysp}; errors are $1\sigma$).
With 90 per cent confidence the temperature exceeds 2.9~keV. The
intrinsic ({\sl ie} corrected for Galactic absorption) 0.3-10\keV\
(rest-frame) luminosity is 4.5$\times10^{44}$\ergps; the 2-10\keV\
luminosity is $(2.5\pm0.4)\times10^{44}$\ergps. Alternatively, if we
fit the extended X-ray emission by a cooling plasma (Johnstone et al
1993) at an abundance 0.4$\times$~Solar (Allen \& Fabian 1998), the
best fit (reduced-$\chi^2$ of 1.44 for 6 d.o.f) gives a gas cooling
rate of $400-700$\Msunpyr from a temperature of $>15$ to 8~keV,
respectively. The 90 per cent lower limit on the cooling flow
temperature is 7~keV. We repeated the spectral fitting of the extended
emission now excluding the region around the SW hotspot and find that
the results do not differ significantly. The spectrum is also
consistent with power-law emission (photon index of $1.96\pm0.35$).

We also examined the spectrum of the point X-ray source associated
with the radio core, but with $\sim30$ counts we can only obtain
approximate properties. Most of the counts from this source are only 
above 2\keV; from modelling the spectrum with a power-law model of
photon index $\Gamma=2$, this implies an excess absorption ({\sl ie}
over the Galactic column) of $\sim7\pm3\times10^{23}$\pcmsq. The total
(de-absorbed) 2-10\keV\ (rest-frame) luminosity of the nucleus is then
$\sim1.1\times10^{45}$\ergps. Both the intrinsic luminosity and
line-of-sight absorption to the central powerhouse makes it comparable
to the central source of Cygnus~A (Ueno et al 1994), and steep
spectrum radio quasars in general.

\section{ Discussion}
\subsection{ Inverse-Compton emission?}
The stronger low-redshift powerful 3C galaxies, such as Cygnus~A
(Wilson, Young \& Shopbell 2000) and 3C295 (Harris et al 2000), show
soft X-ray emission associated with the position of outer radio lobes.
This is interpreted as due to Inverse-Compton (IC) scattering of
cosmic microwave background photons by the relativistic electrons in
the radio plasma. The energy density of the microwave background is 60
times higher at $z=1.786$ than at the current epoch, so the cooling
time of relativistic electrons is short. It has also been predicted
that these electrons can also inverse-Compton scatter photons from the
nucleus, to produce an asymmetric, but spatially extended component of
X-ray emission (e.g. Brunetti, Setti \& Comastri 1997). In this
scenario the X-ray emission arising in the more distant radio lobe is
expected to be brighter due to the stronger back-scattering of photons
towards the observer. We do find tentative X-ray emission associated
with the southern outer hotspot, and also possibly with the inner
hotspot and lobe to this side. There is no obvious excess of emission
associated with the northern radio source components. Assuming that
the southern side is the further lobe, then the observed excess may
fit predictions.

The lack of any clear correspondence of X-ray emission with the
Northern radio hotspot, of any widespread diffuse radio emission, and
the extension of the diffuse X-ray emission beyond the radio source,
indicates that Inverse-Compton scattering by an electron population
related to the radio emission contributes only a small fraction of the
total X-ray luminosity observed with {\sl Chandra}. 

The possibility remains that the emission is due to an older
population of relativistic electrons inverse Compton scattering the
microwave background (see e.g. Sarazin 1999). The required Lorentz
factor is then about 300 and the cooling time of the electrons $t_{\rm
ic}=5\times 10^7\yr$. Can such emission mimic hot cluster gas? We note
that a) to distribute electrons to 100~kpc radius requires supersonic
motion, and b) to confine them requires a substantial atmosphere of
gas with a pressure exceeding that of the relativistic gas (or it
would explode outward). Assuming that the gas pressure required is at
least $a$ times the minimum inferred from the observed X-rays (i.e.
$aL_{\rm X} t_{\rm ic}/V$ where $V$ is the volume within 100~kpc; $a$
must include electron, proton and magnetic pressures), then we find
that the predicted X-ray luminosity of the atmosphere, if
$kT_X=1\keV,$ equals that seen below 1~keV if $a=10$. In other words,
to confine a fossil electron population in a gravitational well
shallower than implied by a thermal interpretation of the X-ray
spectrum overpredicts the observed soft X-ray flux, unless the thermal
overpressure $a<10$.

\subsection{ An intracluster medium}

The spatial scale of the extended soft X-ray component is comparable
with that expected from the inner parts of an intracluster medium
centred on 3C294, particularly if there is a cooling flow in the
cluster (Fabian 1994). This interpretation is supported by its
spectrum; the relatively high temperature of the gas, $kT>2.9\keV$,
indicates that we are dealing with a cluster, and not just the hot
halo of a massive galaxy. The 0.3-2\keV\ luminosity we find for the
extended component from the {\sl Chandra} data
($\sim2\times10^{44}$\ergps) is in good agreement with the 0.7-2\keV\
luminosity of $1.7\times10^{44}$\ergps we inferred from the {\sl
ROSAT} PSPC data. The {\sl Chandra} data confirm that any nuclear
X-ray emission is strongly absorbed below 2\keV, and thus could not
have made a major contribution to the observed {\sl ROSAT} flux from
3C294. We thus confirm our original supposition that the {\sl ROSAT}
X-ray emission associated with this source was from a surrounding
cluster of galaxies (Crawford \& Fabian 1996; see also Hardcastle \&
Worrall 1999; Dickinson et al 1999). The X-ray luminosity observed
suggests that 3C294 is embedded in only a relatively modest cluster,
although at this large a distance we are probably only seeing the
central regions of a cluster with a cooling flow. The cluster has
about half the luminosity of the cluster surrounding the powerful
low-redshift FR~II radio source Cygnus~A. It also fits the
temperature--luminosity function for nearby clusters (Fig.~5).

The asymmetric hourglass structure of the diffuse X-ray emission is
different from that generally seen in low-redshift clusters. The X-ray
data raise the intriguing possibility that the triangular shapes
observed in the Lyman-$\alpha$ and K$'$ images are {\sl not} due to an
illumination cone prescribed by the opening angle for escaping quasar
radiation, but instead arise from the physical distribution of the gas
around the source. If this is the case, it is curious that the regions
apparently deficient in X-ray gas are {\sl not} associated with the
radio source direction, in contradiction to the X-ray cavities seen in
{\sl Chandra} images of clusters around nearby radio sources (McNamara
et al 2000; Fabian et al 2000). If the inner hotspot describes the
current pointing position of the radio jets, they seem to be directed
almost into the densest extended X-ray emission. If the radio outflow
was oriented more E--W at a much earlier stage of the central source
(consistent with the direction of the jet precession) it may well have
shaped the X-ray gas. Of course such cavities need not be devoid of
gas, since the medium may be multiphase and the brighter parts may
just be those where the cooler gas is most abundant.

A peaked, bright X-ray core to a cluster is characteristic of a
cooling flow, such as are seen around powerful radio galaxies such as
Cygnus A (Reynolds \& Fabian 1996) and 3C295 (Allen et al 2001).
Extended Ly-$\alpha$ emission is also seen from the nebulosities in
low redshift cooling flows (Fabian et al 1984; Hu 1988), although the
Ly-$\alpha$ luminosity of 3C294 is very high (and comparable with the
X-ray luminosity). The existence of a cooling flow in the 3C294
cluster is plausible since the mean density of the gas within 100~kpc
of the radio source is $0.02\pcmcu$ and the mean cooling time is about
2~Gyr, less than the age of the Universe at that epoch (3~Gyr, for the
adopted cosmology). The pressure of the intracluster medium is
consistent with the pressure determined from equipartition arguments
for the low surface brightness radio emission around the Southern
radio emission by McCarthy et al (1990). Strong Faraday rotation is
also observed for the radio source (Liu \& Pooley 1991), a further
characteristic of cooling flows (Taylor, Barton \& Ge 1994). Our
results reinforce our earlier hypothesis that powerful radio galaxies
may be a way to discover distant cooling flows (Fabian et al 1986).

As a final consideration on the surrounding gas, we have estimated
whether electron scattering of nuclear X-ray emission could contribute
significantly to the observed flux (and thus the inferred
temperature). Given the parameters of the gas and the central source,
the Thomson depth is $0.004$ and the scattered flux (assuming that
half the Sky is obscured at the nucleus) is only about one per cent of
the observed flux. If instead the flux detected from the nucleus is
also scattered, so that the nucleus is really much more luminous, then
we require its 2--10~keV luminosity to be about 2 orders of magnitude
greater. (The Thomson depth hardly changes if we reduce the gas
temperature to 1~keV and find the densest cluster consistent with the
spectrum.) Its bolometric luminosity is then $\sim 5\times
10^{48}\ergps$, assuming that the spectral energy distribution follows
that for quasars found by Elvis et al (1994). Much of this should be
absorbed and reradiated in the far infrared. It is ruled out by 60 and
100~$\mu$m limits from IRAS, which are about an order of magnitude
smaller ($1\sigma$ levels of 28~mJy and 111mJy at 60 and 100$\mu$m
respectively have been obtained using the web-based XSCANPI at IPAC).
We conclude that scattered X-ray emission is unimportant.

\subsection{ The occurrence of such high-redshift clusters}

Following the calculation of Donahue \etal (1998) for the cluster
MS~1054--0321, we now estimate how rare a cluster resembling that
around 3C294 might be at this redshift. For simplicity we first adopt
an $\Omega_0=1$ universe, and use the results from our MEKAL spectral
fit, with a conservative lower limit to the temperature of the
intracluster medium of 3\keV. The temperature of an equivalent cluster
at $z=0$ is thus $kT_X>1$\keV\ ($T_X\propto(1+z)$ for $\Omega_0=1$),
and its virial mass is greater than $1.3\times10^{14}$\Msun (Henry
2000). We extrapolate the temperature function of Henry (1997) to
estimate that the present-day number density of clusters hotter than
1\keV\ is less than $2.4\times10^{-5}$\pMpccu. Thus the mean
virialized mass density in such clusters is less than
$3.3\times10^9$\MsunpMpccu. The assumption of Gaussian perturbations
in an $\Omega_0=1$ universe allows us to use the integral form of the
Press-Schechter (1974) formula to derive the comoving mass density of
virialized objects with virial masses greater than $M$ from the
current matter density, $\rho_0$, and $\nu_c$ is the critical
threshold at which the perturbations leading to these structures
arise. For a present-day cluster with $kT_X>1$\keV, $\nu_c>2.0$. As
$\nu_c\propto(1+z)$ for a fixed mass scale, this implies that
$\nu_c>5.6$ for similar clusters at $z=1.786$ ({\sl ie} $kT_X>3$\keV\ 
then). Using the Press-Schechter formula to obtain the comoving
virialized density of such systems as a function of redshift, we
integrate the result for $z>1.786$, assuming the area of the 3C
catalogue (north of $\delta=-5$\degmark), to obtain the predicted
total number of clusters (Fig.~6, which also shows results for other
values of $\Omega_{\rm m}$). The detection of a $kT_X>3\keV$ cluster
at $z=1.786$ is inconsistent with $\Omega=1$.

\begin{figure}
\psfig{figure=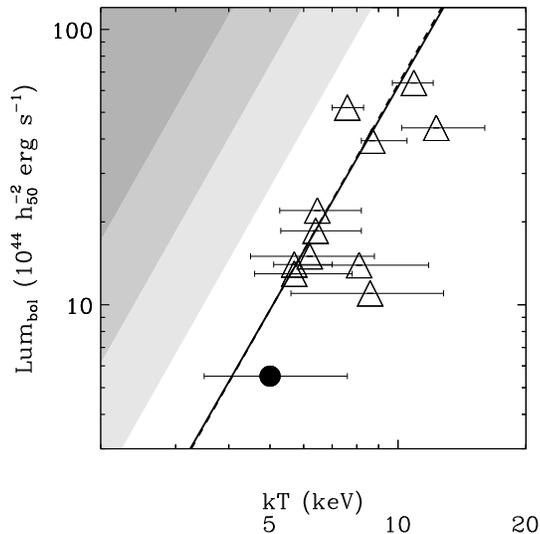,width=.45\textwidth}
\caption{\label{fig:tl}
Luminosity -- temperature relation for galaxy clusters with $z>0.5$
(EMSS; Donahue et al 1999, RDCS Della Ceca et al 2000, Schindler
1999). The dot represents 3C294. The solid line is the best-fit for
data of a sample of galaxy clusters with temperature and luminosity
corrected by the presence of cooling flows (Ettori, Allen, Fabian
2000). The dashed line is the best-fit for nearby clusters from Wu,
Xue and Fang (1999). The shaded region on the left show the expected
shift of the $L-T$ relation for evolution with $1<A<2$, $2<A<3$ and
$A>3$ at $z=1.786$ using the relation $L \sim T^s (1+z)^A$. As shown
by Borgani et al. (1999), $\Omega_0=1$ models require positive
evolution of the $L-T$ relation (i.e. $1<A<3$), whereas no evolution
implies low $\Omega_0$ cosmologies. }
\end{figure}

\begin{figure}
\psfig{figure=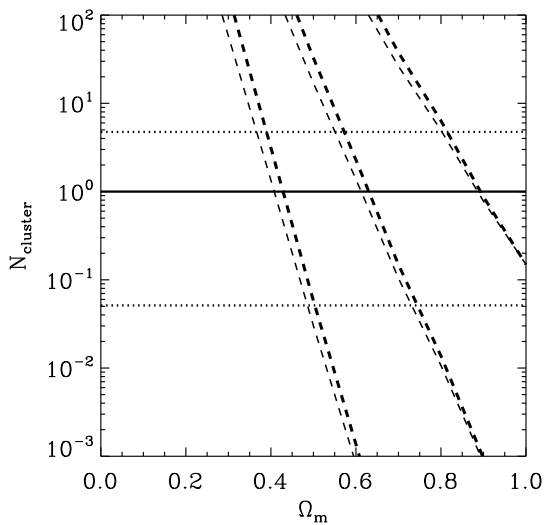,width=.45\textwidth}
\caption{ \label{fig:prob}
{\it Maximum} number of collapsed objects beyond
redshift 1.786 with virial mass larger than
that corresponding to 3, 5 and 9 keV (from right to left). The dashed
lines are for $\Omega_{\rm m} + \Omega_{\Lambda}=1$ (thick) and
$\Omega_{\Lambda}=0$ (thin). The single detection puts an upper limit
on $\Omega_{\rm m}$. The horizontal lines represent 90 per cent
uncertainties on a single detection (Gehrels 1986).}
\end{figure}

\section{Conclusions}
The {\sl Chandra} observation of 3C294 reveals soft, spatially
extended, X-ray emission that is clearly resolved from any X-ray
emission associated with the radio source components. We detect the
radio nucleus as a heavily obscured point source, and a possible
excess of X-ray emission at the site of the southern radio hotspots.
The diffuse X-ray emission has an unusual hour-glass morphology that
is roughly aligned near the centre with the extended Lyman-$\alpha$
and K$'$ emission associated with the radio galaxy. The soft extended
component is centred on the active nucleus, and with a diameter of
over 200\kpc\ is on a larger scale than that of the embedded radio
source. Its X-ray spectrum is fitted well by a thermal model with
temperature of at least 3\keV, and we identify it as the intracluster
medium around 3C294.  At $z=1.786$, this is the first X-ray detection
of a cluster of galaxies above $z=1.3$. Its temperature and the lack
of evolution in the $L_X-T_X$ relation out to this redshift both
support $\Omega<1$.

Further, much deeper, observations with {\sl Chandra} will enable both
the radio and diffuse hot gas components to be better studied. Of
great interest for the latter are the temperature and density
structure of the intracluster medium, as well as its metallicity.
$\Omega_{\rm m}$ can also be constrained from the observed temperature.

\section{Acknowledgements}
We are grateful to NASA and the {\sl Chandra} project for the superb
X-ray data and thank Steve Allen for discussions. ACF and CSC thank
the Royal Society for financial support.

{}

\onecolumn

\end{document}